# DYNAMICS AND STRUCTURE OF PMN AND PZN

G. Shirane and Guangyong Xu

Physics Department, Brookhaven National Laboratory, Upton, NY 11973, USA

P. M. Gehring

National Institute of Standards and Technology, NIST Center for Neutron Research, Gaithersburg, MD 20899, USA

ABSTRACT

A review is given of recent neutron and x−ray scattering studies of the lead−oxide perovskite relaxor systems $Pb(Zn_{1/3}Nb_{2/3})O_3-xPbTiO_3$ and $Pb(Mg_{1/3}Nb_{2/3})O_3-xPbTiO_3$. X−ray measurements by Noheda *et al.* have established that these two systems exhibit nearly identical phase diagrams in which a rhombohedral−monoclinic−tetragonal structural sequence takes place with increasing $PbTiO_3$ concentration. Recent high−energy x−ray and neutron measurements on single crystals of PZN and PMN−10PT, however, show that the rhombohedral distortions occur only in the outermost 20 − 40 microns, while the bulk of each crystal transforms into a new phase X, which has a nearly cubic unit cell. This situation is very similar to the structural behavior of pure PMN at $T_c = 220$ K. A simple model has been suggested that correlates phase X to the unique atomic displacements of the polar nanoregions, which are created at the Burns temperature.

INTRODUCTION

There are two important aspects to the structural characterization of the relaxors Pb(Zn$_{1/3}$Nb$_{2/3}$)O$_3$ (PZN), Pb(Mg$_{1/3}$Nb$_{2/3}$)O$_3$ (PMN), and their solid solution with PbTiO$_3$ (see Ref. [1]). One is the basic phase diagram plotted as a function of PbTiO$_3$ (Ti$^{4+}$) concentration. The other is the structural modulation created by the appearance of the polar nanoregions at the Burns temperature $T_d$. We demonstrate in this review that these two aspects are intimately coupled.

MONOCLINIC STRUCTURE

In 1999 Noheda *et al*. [2,3] discovered a new monoclinic structure in PbZr$_{1-x}$Ti$_x$O$_3$ (PZT) located between the rhombohedral and tetragonal phases, just to the left (lower $x$) of the morphotropic phase boundary near $x = 50\%$. This surprising finding provided a new perspective for the structural concept in ferroelectric oxides. Similar monoclinic phases were discovered soon thereafter in the relaxor systems PZN−$x$PT [4] as well as PMN−$x$PT [5] as shown in Fig.1.

The monoclinic structure has the unique property that allows the polar direction to rotate within the basal plane; this freedom stands in sharp contrast to the uniaxial constraint imposed on the polar direction in both rhombohedral and tetragonal symmetries. With this property in mind, the two monoclinic structures, $M_A$ and $M_C$, can be understood as resulting from, respectively, the two different polarization paths [7] as shown in Fig. 1d and Fig. 1e.

Vanderbilt and Cohen [8] presented an elegant way to describe the relationship between these new structures with the old one. By including the next highest term in the Devonshire free energy expansion, a new global phase diagram was created. This simple extension of the Devonshire theory admits all three kinds ($M_A$, $M_B$, $M_C$) of monoclinic phases. In their study,

Vanderbilt and Cohen emphasized that these monoclinic phases can only be realized in extremely anharmonic solids. Then the free energies of the cubic, tetragonal, rhombohedral, and monoclinic phases lie very close to each other.

This special feature manifests itself in various relaxor properties, and may be responsible for their ultrahigh piezoelectric response. It is also reflected in some of the scattering profiles. One example of this is shown in Fig. 2 where several different Bragg reflections of a powder sample of PMN−35PT were studied with x−rays as a function of temperature [5]. This particular composition goes through a cubic−tetragonal−monoclinic phase sequence on cooling. As one can see clearly in Fig. 2, the x−ray profiles in the cubic (450 K) and tetragonal (400 K) phases are very sharp. Only in the monoclinic phase do the profiles become very broad. This is not due to a co−existence of two phases in the usual sense. Rather it is the flatness of the free−energy surface that permits two monoclinic structures, with slightly different lattice parameters, to coexist. In a similar fashion, an optical study of a crystal with a similar $PbTiO_3$ concentration reported finding different polarization directions in different parts of the crystal [9].

POLAR NANOREGIONS

Perhaps the seminal property of relaxors is the formation of nanometer−sized regions of randomly oriented local polarization, or polar nanoregions (PNR). Measurements of the optic index of refraction on a variety of disordered ferroelectric compounds, including both PZN and PMN, by Burns and Dacol provided the first evidence of these regions which first appear several hundred degrees Kelvin above the respective Curie temperatures [10]. This temperature has since become known at the Burns temperature $T_d$. Naberezhnov *et al*. demonstrated that neutron diffuse scattering appears in PMN between 600 and 650 K [11], consistent with the optical measurements of Burns and Dacol. For this reason the diffuse scattering that develops in relaxors below $T_d$ has been identified with the PNR. Subsequent measurements by Stock *et al.* have shown that the same identification holds in PZN as well [12].

The PNR also have a strong influence on the dynamical properties of relaxors, particularly in regards to the lattice dynamics. Early neutron inelastic scattering studies by Gehring *et al.* on PZN−8PT, PZN, and PMN, showed that the transverse optic (TO) phonon energy linewidths are strongly damped in the cubic phase for wavevectors $q$ close to the zone center [13−15]. Because the phonon damping is highly $q$−dependent, constant−intensity contour plots of the inelastic scattering exhibit an unusual feature whereby the TO phonon branch appears to drop sharply into the TA phonon branch at non−zero $q$. This anomalous feature was dubbed the "waterfall" effect. The fact that the phonon damping becomes significant only for sufficiently small $q$ suggests that the PNR inhibit the propagation of long−wavelength phonons when the phonon wavelength is comparable to the size of the PNR. In this regard the waterfall feature provides a measure of the average size of the PNR.

Gehring *et al.* provided a key insight into the origin of the PNR in their high−temperature neutron scattering lattice dynamical study of PMN [16]. Whereas only damped or overdamped TO phonons were observed at the zone center below $T_d$, consistent with earlier studies, a clear zone center TO phonon was observed for the first time at 1100 K (far above $T_d$) that softens and broadens in energy with decreasing temperature. This zone center phonon, which exhibits a temperature−dependence consistent with that of a classic ferroelectric soft mode, becomes overdamped at the Burns temperature, thereby suggesting an intimate connection between it and the formation of the PNR at $T_d$.

Wakimoto *et al*. extended the lattice dynamical study of PMN to lower temperatures [17]. Remarkably, below the PMN Curie temperature $T_c = 213$ K (defined after cooling in a weak applied field) the zone center soft TO mode reappears, resuming a linear dependence of the phonon energy squared with temperature, just as is observed in ferroelectric $PbTiO_3$. These results are summarized in Fig. 3, along with the observation of a significant broadening of the transverse acoustic (TA) mode, and provide a dramatic illustration of the effects of the two temperature scales $T_c$ and $T_d$, on the lattice dynamics of PMN. The softening and then eventual overdamping of the zone center TO mode at the Burns temperature, along with the simultaneous appearance of diffuse scattering, lend strong support to the view that the PNR result from the condensation of this soft mode. At low temperatures the reappearance of the soft mode at $T_c$ suggests the establishment of a ferroelectric phase in PMN, in spite of the absence of any structural distortion.

Measurements on PMN, PZN and other compounds have shown that the diffuse scattering that first appears at $T_d$ continues to grow monotonically with decreasing temperature, even below $T_c$. This is demonstrated in Fig. 4 which shows neutron diffuse scattering intensities measured at $(3,q,q)$ as a function of temperature at several different values of $q$ on a single crystal of PMN−10PT [18]. The data show that, even below $T_c \sim 285$ K, the diffuse scattering continues to increase in strength, thus indicating that the PNR are still present. The inset of Fig. 4 shows a color contour map of the diffuse scattering measured on pure PMN at (100) measured using neutron time−of−flight techniques at 300 K [19].

Hirota *et al*. made a fundamental and critical discovery [20] when they re−examined the ionic displacements, determined at 300 K in PMN by pioneering neutron diffuse scattering measurements reported by Vakhrushev *et al*. [21]. They realized that these displacements could be decomposed into a center−of−mass conserving component that could explain the observed phonon scattering intensities of the TO mode in PMN in different Brillouin zones, and an overall constant scalar shift common to each atom. In this picture the PNR would result from the condensation of the soft mode, but a uniform displacement δ would appear at the Burns temperature as shown schematically in Fig. 5. The shifted or displaced character of the PNR are believed to be a unique feature of these relaxor systems that govern most of their low−temperature properties.

THE NEW PHASE X

Closely related to the questions about the PNR is the discovery of the new phase X in PMN and PZN. The previously accepted zero−field phase diagram for relaxor ferroelectrics indicates that both PMN−$x$PT and PZN−$x$PT remain rhombohedral for PT concentrations $x$ to the left of the MPB and T < $T_c$ [4,5,22−24]. An exception, however, appears to be pure PMN, which retains a cubic structure down to 5 K [25−27].

Several studies have recently uncovered exciting evidence of a new phase on the rhombohedral side of the phase diagram. Studies of the effects of an applied electric field on PZN−8PT by Ohwada *et al*. [30] show that the unpoled system does not have rhombohedral symmetry below $T_c$. This was the first hint that a new phase may be present. The first conclusive evidence for this new phase, which was labeled phase X by the authors, was provided by high−energy x−ray diffraction studies on PZN single crystals by Xu *et al*. [31]. The studies showed that x−ray measurements on single crystals of PZN gave different results for different x−ray energies. Fig. 6 shows longitudinal scans along the pseudocubic (111) direction of an unpoled PZN single crystal with different x−ray energies at room temperature (below $T_c$). When probed with 32 keV x−rays, which penetrate only ∼ 50 μm into the sample, the pseudocubic (111) reflection splits into two peaks. Both peaks have widths slightly larger than the resolution width. If this splitting is related to a rhombohedral distortion, and the two peaks are indeed the (111) and (−111) reflections arising from different domains, then the rhombohedral angle α can be determined from the difference of peak positions. The result ($\alpha = 89.916°$) is in agreement with previous measurements [24].

However a truly remarkable result is revealed by the scan with 67 keV x−rays, with a penetration depth of several hundred microns, which shows only one sharp, resolution−limited peak at (111). No rhombohedral splitting is observed. This result demonstrates unambiguously that the bulk structure of PZN is different from that of the ∼ 50 μm thick outer−layer. While the outer−layer exhibits a rhombohedral structure as previously believed, the bulk does not show any macroscopic rhombohedral distortion. Instead, it has an average cubic unit cell structure ($a \sim b \sim c$). The corresponding symmetry of this new phase X is still unknown.

In contrast to the high−energy x−ray case, considerable line broadening was observed in the neutron profiles on PZN below $T_c$ by Stock *et al*. [12]. This observation was attributed to the fact that neutrons scatter from a far greater volume of the crystal than do x−rays. The current conjecture is that the additional strains that produce this broadening are created by the interaction between the polar nanoregions and the surrounding polar regions.

Motivated by these findings, high *q*−resolution neutron scattering measurements were performed on PMN−10PT by Gehring *et al*. [18]. The results are compared with the earlier x−ray measurements of Dkhil *et al*. on the same compound in Fig. 7. The upper panel shows data obtained using 8.9 keV Cu K$_\alpha$ x–rays. A clear splitting of the (222) Bragg peak occurs below T$_c$. The lower panel shows the neutron data, which probe the bulk of the crystal, and which reveal a single peak at (111) reflection for temperatures both above and below T$_c$, and thus no rhombohedral distortion, thus directly contradicting the x–ray result.

These data demonstrate that the rhombohedral phase previously believed to be present in these relaxor ferroelectric systems below T$_c$ is actually phase X. Pure PMN is most likely phase X too, but could represent a limiting case of a very thin outer layer. In fact, more recent neutron scattering measurements on PMN–20PT and 27PT single crystals show that the bulk of the 20PT exhibits phase X, while 27PT transforms into a rhombohedral phase below T$_c$ [32]. An important question concerns the origin of phase X, and its relationship with the polar nanoregions. We believe that the uniform phase shift [20] of the PNR plays an essential role in stabilizing phase X by creating an energy barrier that prevents the PNR from melting into the surrounding ferroelectric phase. The balance between this energy barrier and the coupling between the polar nanoregions and the surrounding environment determines the stable phase of the system below T$_c$. With higher PT concentrations, the coupling becomes stronger and eventually phase X develops into the rhombohedral phase in PMN–xPT. A new and universal phase diagram that reflects this concept is shown in Fig. 8.

DISCUSSION

We have reported several new observations on the structural aspects of PMN, PZN, and their solid solutions with PbTiO$_3$. There are many important unanswered questions left, such as the temperature dependence of the size and density of the polar nanoregions though T$_c$. It has been conjectured that the shifted, or displaced, nature of the PNR, as described in earlier has a profound influence on the stability of the new phase X below $T_c$.

One of the key experiments needed to test this conjecture is the study of the effect of an electric field on the diffuse scattering. Structural changes of PZN−8PT under electric filed were reported previously [28,29]. The large jump of the lattice constant around 15 kV/cm, at room temperature, was explained by the $M_C$ to Tetragonal phase transition. Recently, a more systematic study of an [001]−oriented electric field on PZN−8PT was reported by Ohwada *et al.* [30], the results of which are illustrated in Fig. 9. A very small electric field of 200 V/cm is sufficient to create a single−phase state below $T_c$ (500 K). Under these conditions the system first transforms from a cubic to tetragonal phase, and then to the $M_C$ phase, on cooling. On the other hand, when cooled in zero field (ZFC) the system becomes tetragonal as before, but then transforms to phase X, instead of $M_C$. The $M_A$ phase is realized only by the specific route of ZFC to below 400 K, and then applying an [001]−oriented electric field. Preliminary results of the diffuse scattering indicate that the diffuse scattering intensity around the (003) Bragg peak is considerably reduced at all temperature below $T_c$, while the diffuse scattering around (300) remains strong. Further study along this line, as well as the [111]−oriented field case, will be reported shortly.

Finally, what is the relationship between the monoclinic phases, phase X, and the polar nanoregions below $T_c$? We have not yet found a simple model that relates all of these features. However, the highly anharmonic oxide model proposed by Vanderbilt and Cohen suggests that all of these unusual phenomena are the result of an extremely flat free energy surface. Thus the stabilities of various phases lie so close together in energy that the domain boundaries, such as the shifted PNR, could create unexpected structural geometries. Phase X appears to be an extreme example of this in which the usual coupling between the polar order parameter and the expected lattice distortion is broken.

ACKNOWLEDGMENTS

We would like to thank our collaborators for many stimulating discussions. We wish to pay special tribute to the late Seung–Eek "Eagle" Park. His passion for research and his kind nature made our joint collaborations both productive and enriching. Work at Brookhaven National Laboratory is supported by the U.S. Department of Energy under Contract No. DE–AC02–98CH10886. We also acknowledge the support of the NIST Center for Neutron Research, the U.S. Department of Commerce, for providing the neutron facilities in this work.

FIGURES

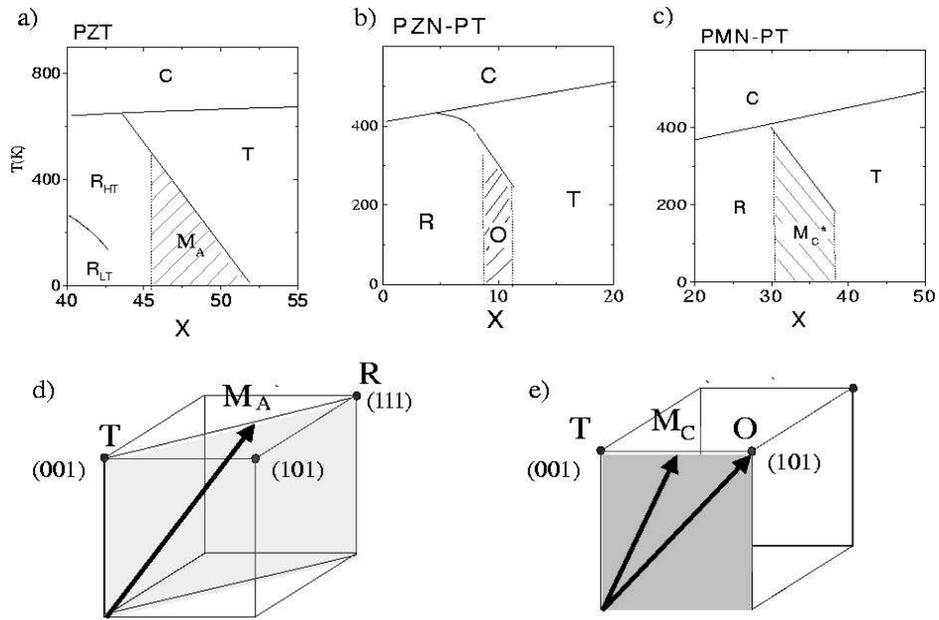

Fig. 1. Phase diagrams are shown for a) PZT, b) PZN−PT, and c) PMN−PT. The hatched regions indicate the $M_A$ and $M_C$ phases located near the MPB. From Noheda [6].

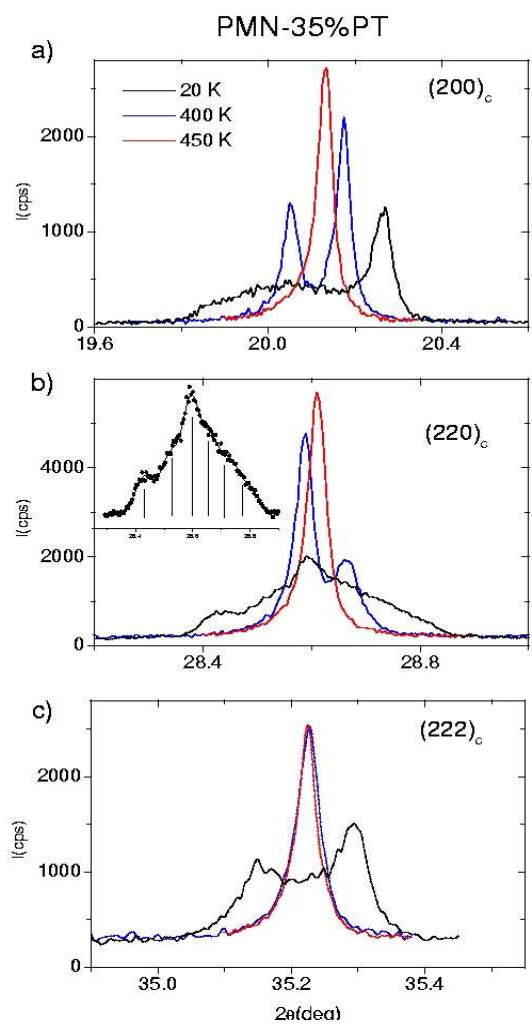

Fig. 2. X−ray scattering profiles of the (200) (220), and (222) Bragg reflections measured at 20, 400, and 450 K on a powder sample of PMN−35%PT. From Noheda *et al*. [5].

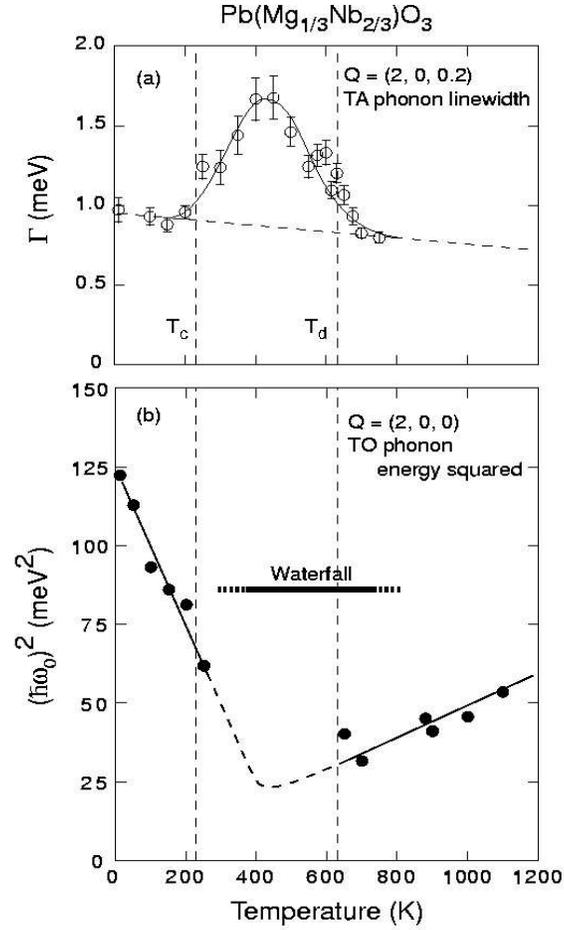

Fig. 3. (a) Plot of the intrinsic TA phonon linewidth $\Gamma$ (open circles) versus temperature measured at $\mathbf{Q} = (2,0,0.2)$. Solid line is a guide to the eye. (b) Plot of the square of the zone center TO phonon energy versus temperature measured at $\mathbf{Q} = (2,0,0)$. The solid line is a fit to a linear temperature dependence, which is the hallmark of a ferroelectric soft mode. From Wakimoto *et al*. [17].

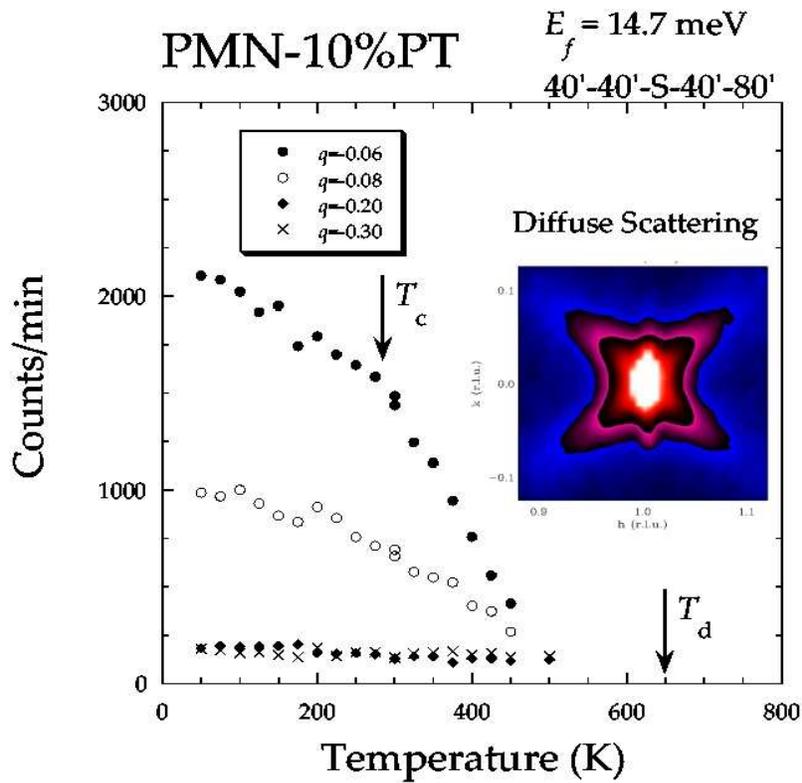

Fig. 4. Temperature dependence of the neutron diffuse scattering measured at **Q** = (3,*q*,*q*) on a single crystal specimen of PMN−10PT. From Gehring *et al*. [18]. The inset shows a color contour map of the diffuse scattering around **Q** = (1,0,0) in pure PMN as measured on the NIST Center for Neutron Research Disk Chopper Spectrometer at 300 K. From Xu *et al*. [19].

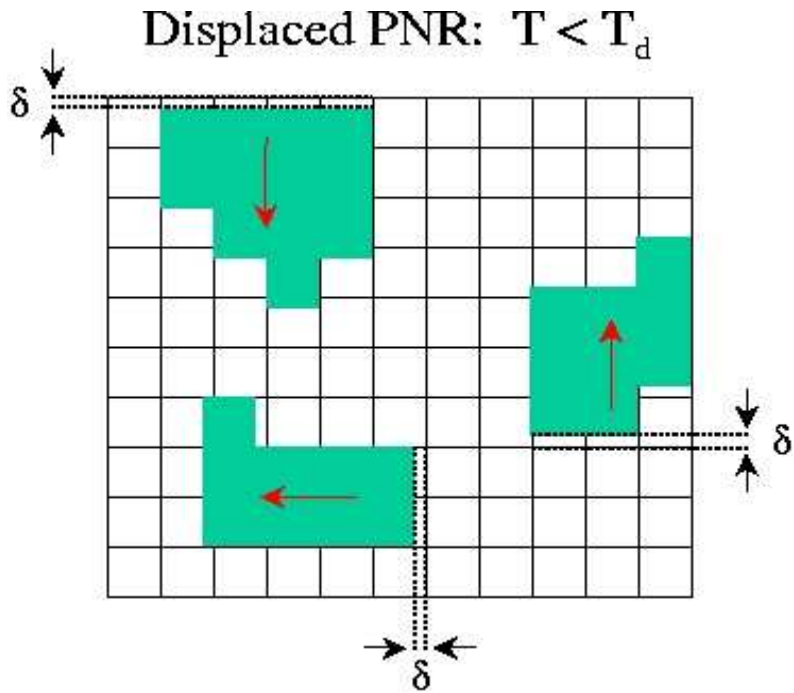

Fig. 5. A schematic representation of the polar nanoregions (shaded) surrounded by the undistorted cubic (white) lattice. The shift δ is the same for all atoms within a given polar nanoregion, and is measured with respect to the cubic lattice.

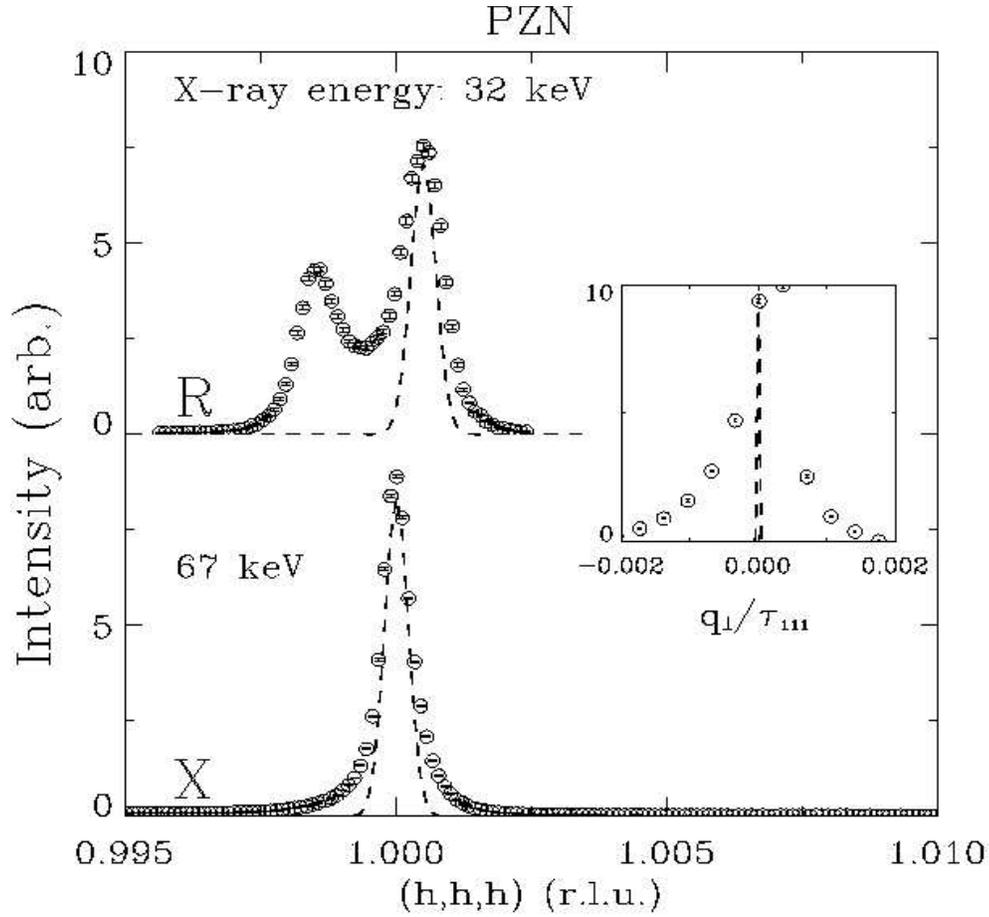

Fig. 6. Longitudinal scans (θ−2θ scans) of the (111) Bragg reflection measured on an unpoled PZN single crystal using different x−ray incident energies. The instrumental resolution was measured using the (111) reflection of a perfect Ge crystal, and are shown by the dashed lines. The inset is a transverse scan through (111) taken at an x−ray energy of 67 keV. The horizontal axis is plotted in units of the pseudocubic reciprocal lattice vector $\tau_{111} = ?3\,(2\pi)/a_0$, where $a_0$ is the unit cell lattice spacing. From Xu *et al*. [31].

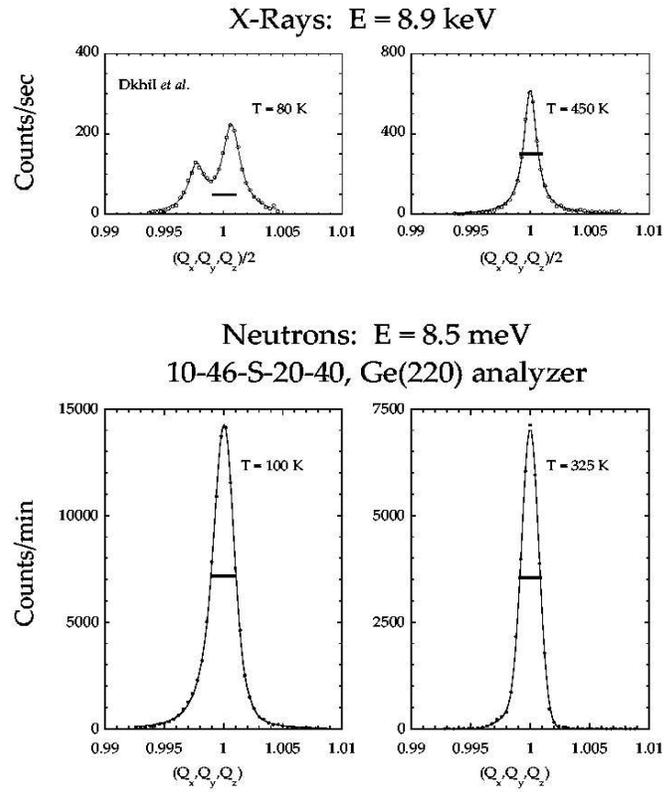

Fig. 7. Top panel: X−ray measurements of the (222) Bragg reflection in PMN−10PT by Dkhil et al. at 80 and 450 K [10]. Bottom panel: Neutron measurements of the (111) Bragg peak in single crystal PMN−10PT by Gehring et al. at 100 and 325 K [18]. The solid horizontal bar in each panel indicates the neutron spectrometer instrumental $q$−resolution FWHM.

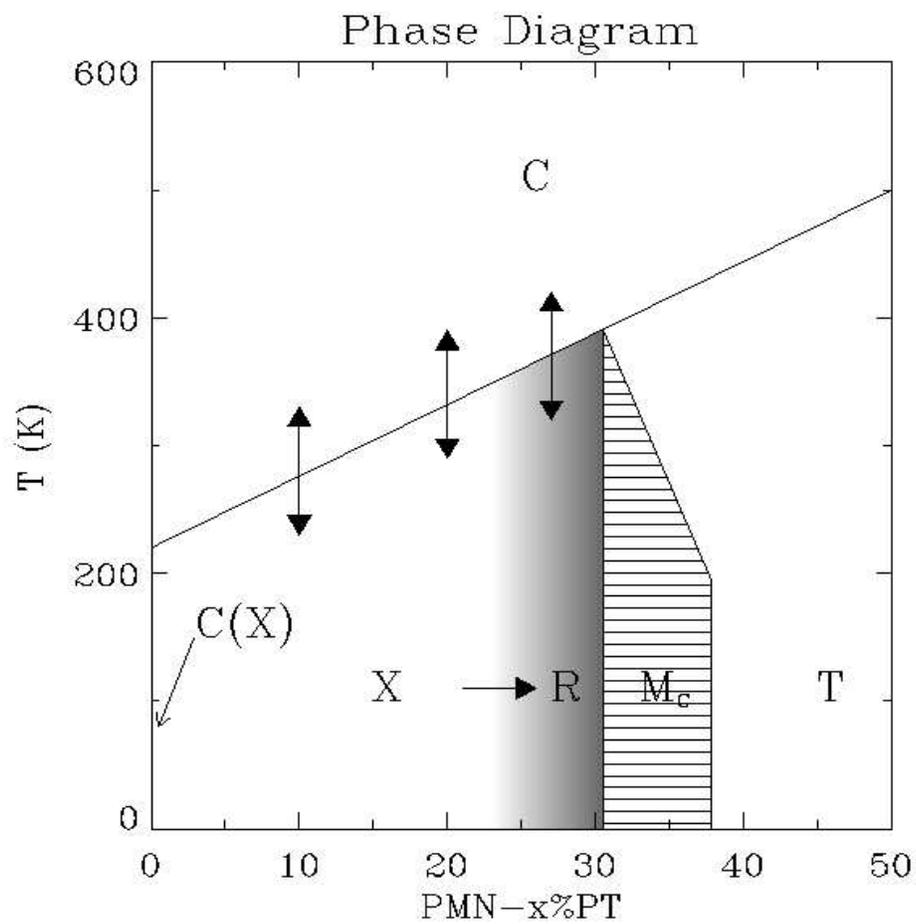

Fig. 8. The revised zero–field phase diagram of PMN–$x$PT. From Xu *et al*. [32].

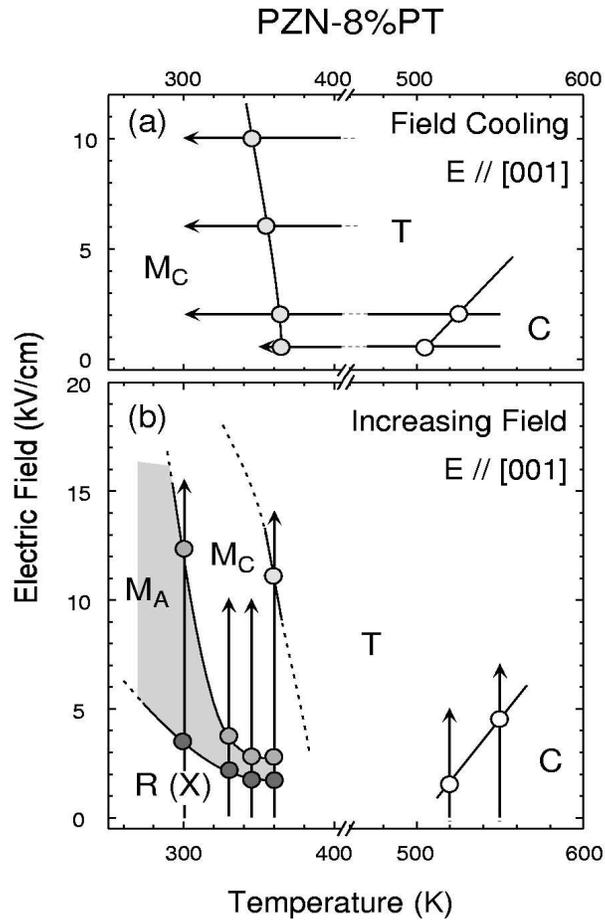

Fig. 9. Electric field versus temperature phase diagram for PZN−8PT. Data in (a) were obtained from field−cooled structural measurements. Data in (b) were obtained on increasing the electric field after ZFC (and ZFH for phase X). The arrows indicate both scan directions, and the ranges of the corresponding measurement sequences. Circles represent the transition temperatures and fields determined from each sequence. From Ohwada *et al.* [30].